\input harvmac
\baselineskip=.55truecm
\Title{\vbox{\hbox{HUTP--97/A001}\hbox{hep-th/9701015}}}
{\vbox{\centerline{N=1 Dualities of  SO and USp Gauge Theories}
\vskip .1in
\centerline{and T-Duality of String Theory}}}
\vskip .1in
\centerline{ Cumrun Vafa\foot{E-mail:
vafa@string.harvard.edu}
and Barton Zwiebach\foot{E-mail:zwiebach@irene.mit.edu}}
\vskip .2in
\centerline{\it Lyman Laboratory of Physics, Harvard University}
\centerline{\it Cambridge, MA 02138, USA}


\vskip .2in
\centerline{ABSTRACT}
\vskip .2in
Extending recent work on $SU$ gauge theory,
we  engineer local string models for $N=1$
four-dimensional $SO$ and $USp$ gauge theories
coupled to matter in the fundamental.
The local models are type IIB orientifolds with
D7 branes on a curved orientifold 7-plane,
and matter realized by adding D3 branes on
the orientifold plane.  The Higgs branches of the
$SO$ and $USp$ theories can be matched
with the moduli spaces of $SO$ and $USp$
instantons on the compact
four-dimensional part of the D7 branes worldvolume.
The R-charge of the gauge theories
is identified with a $U(1)$ symmetry on the
worldvolume of an Euclidean D3 brane instanton.
We argue that the quantum field theory
dualities of these gauge theories arise from T-dualities of type IIB
strings exchanging D7 and D3 charges.
A crucial role is played by the induced D3 charge
of D7 branes and an orientifold 7-plane, both partially compactified on
a $Z_2$ orbifold of $K3$.
\Date{\bf January 1997}

\newsec{Introduction}

It has become increasingly clear over the last year that
many results in supersymmetric gauge field theories
can be derived rather efficiently using string theory.
In particular  $T$-duality symmetries
of perturbative strings has emerged as an extremely powerful tool
in this connection.
For the case of $N=4$ supersymmetric field theories strong-weak coupling
duality is argued to arise from $T$-duality\ref\wittensvd{
E. Witten, `String theory dynamics in various dimensions', Nucl. Phys.
{\bf B443} (1995) 85; hep-th/9503124.}\ref\duff{M. Duff, `Strong/weak
coupling duality from the dual string', Nucl. Phys. {\bf B442} (1995) 47;
hep-th/9501030.}\foot{The original
discussions were in the context of string-string dualities, but
such non-trivial string dualities
are not strictly necessary for this connection.
 A local model can be constructed
as type IIA strings in the background
of ALE spaces times a two-torus. A T-duality transformation
of the two-torus maps to an S-duality of the $N=4$ Yang-Mills theory.
The only non-trivial ingredient in this argument, is the fact that ALE spaces
give rise
to non-perturbative enhancement of gauge symmetry through
wrapped D2 branes.}. The non-perturbative
physics of $N=2$
supersymmetric field theories can be derived from
mirror symmetry \ref\katzklemmvafa{S. Katz, A. Klemm, and
C. Vafa, `Geometric engineering of quantum field theories' hep-th/9609239.}, a
symmetry that amounts to $T$-duality of type II strings
\ref\stromingeryau{A. Strominger, S.T. Yau, and E. Zaslow, `Mirror symmetry
is T-duality', Nucl. Phys. {\bf B479} (1996) 243;
hep-th/9606040.}\ref\morrison{D. Morrison, `The geometry underlying
mirror symmetry', alg-geom/9608006.}\ref\mgro{M. Gross, `Mirror symmetry
via 3-tori for a class of Calabi-Yau threefolds', alg-geom/9608004.}.
In many of these works the basic idea
is building a local string model for the gauge field theory.
This is done by isolating
the part of the compactification data which is
relevant for field theory questions and taking the limit where gravity
is turned off.  In the resulting local model the compactification data
is replaced by the relevant non-compact piece
of the internal space.

$N=1$ supersymmetric theories and their dualities, have also
begun to be understood in the context of string theories.
In particular, $N=1$ pure Yang-Mills
was engineered in  \ref\katzvafa{S. Katz and C. Vafa,
`Geometric engineering of $N=1$ quantum field theories', hep-th/9611090.}\
in the context of F-theory compactification on
elliptically fibered Calabi-Yau fourfolds.
This amounts to considering a space $S$ of complex dimension two (a
codimension one subspace of the base), over
which the elliptic fibration acquires an ADE singularity.  Moreover
it was shown that (to avoid adjoint matter)
the space $S$ must satisfy $h^{2,0}(S)=h^{1,0}(S)=0$.
More recently, and  for the case of $SU(N_c)$,
this was extended to include $N_f$ fundamentals and anti-fundamentals
\ref\bjpsv{M. Bershadsky, A. Johansen,
T. Pantev, V. Sadov and C. Vafa, `F-theory, Geometric Engineering and
$N=1$ Dualities', hep-th/9612052.}, by adding $N_f$ D3 branes filling
four-dimensional spacetime and bringing them close to $S$.
 It was checked that this
 local string model does indeed reproduce some of the well known results
concerning these gauge theories. Higgs branches of these gauge theories
were correctly identified with the  moduli spaces of $SU(N_c)$ instantons
on $S$.
Nonperturbative generation of superpotentials, and quantum corrections
to moduli spaces of the field theories were seen to arise when expected
through the effects of euclidean three branes in the local models.
Moreover, it was argued that both the quantum field theory dualities between
$N=2$ supersymmetric $SU(N_c)$ gauge theory with $N_f$ flavors and
$SU(N_f-N_c)$ gauge
theory with $N_f$ flavors \ref\antoniadis{I. Antoniadis and B. Pioline,
`Higgs branch, hyperkahler quotient and duality in SUSY $N=2$ Yang
Mills Theories' , hep-th/9607058.}, and the
corresponding $N=1$ dualities, that hold upon addition
of some neutral matter \ref\seiberg{N. Seiberg, `Electric-magnetic duality
in supersymmetric non-abelian gauge theories', Nucl. Phys.
{\bf B 435} (1995) 129;
hep-th/9411149.}, arise
in the string models as $T$-duality transformations of type IIB.
This makes sense even though one is discussing F-theory
vacua which generically have no
T-duality symmetries. This is
 because in the context of $SU$ gauge groups one can realize
the F-theory local model
by D7 branes of a fixed perturbative type IIB string
\ref\vft{C. Vafa, `Evidence for F-theory', Nucl. Phys. {\bf B469} (1996) 403;
hep-th/9602022.}\
for which the T-duality symmetry applies.
The T-duality transformation roughly speaking inverts
the volume of $S$.  This, in particular, exchanges
D3 and D7 brane charges.  Due to an induced D3 brane
charge on the curved D7 worldvolume, this symmetry ends up
exchanging $N_c\leftrightarrow N_f-N_c$ instead of $N_c\leftrightarrow N_f$
\bjpsv .

It is natural to ask if the results of \bjpsv\
can be extended to other groups and in particular to $SO$ and
$USp$ groups.   In fact a similar story should repeat at least
for $SO$, because F-theory backgrounds with $SO$ gauge symmetry
can be viewed as type IIB orientifolds  \ref\sen{A. Sen,
`F-theory and Orientifolds', Nucl. Phys. {\bf B475} (1996) 562;
hep-th/9605150.} for which T-dualities apply and
can shed light on $N=1$ dualities.
 Similarly, since the difference between
the $SO$ and $USp$ theories at the string perturbative level is the choice
of the sign for the diagrams with odd number of
crosscaps (as we will review in section 2) the arguments
work with equal ease for the $USp$ theories.
In the present paper we will extend the results of \bjpsv\ to the
case of $N=1$, four dimensional
 $SO(N_c)$ gauge field theories
with $2N_f$ flavors of matter
in the fundamental, and to the case of $USp(N_c)$ gauge theory with
$2N_f$ flavors in the fundamental.\foot{In our conventions $USp(N)$,
with $N$ necessarily even, is
the compact gauge group defined by the set of $N\times N$ complex
matrices that are both unitary ( belong to $U(N, {\bf C})$) and
symplectic (belong to $Sp (N, {\bf C})$). The corresponding Lie algebra
is the real Lie algebra $usp(N)$ defined by the set of $N\times N$
antihermitian matrices $A$ that in addition satisfy
the symplectic condition $JA = -A^t J$
with $J= \pmatrix{0& I_{N/2}\cr -I_{N/2} & 0 \cr}$.
The $usp(N)$ algebra is of rank $N/2$, and real dimension
$N(N+1)/2$. The fundamental representation of this algebra is of
dimension $N$ and it is pseudoreal,
and the adjoint representation appears in the symmetric
part of the product of two fundamentals.}  In particular
we show that the Higgs moduli is mapped to the instanton moduli
space on $S$.  Moreover we argue that the gauge symmetry
duality is mapped to T-duality which exchanges $SO(N_c)$ with
$SO(2N_f-N_c+4)$.  The dual group differs from the
naive expectation $SO(2N_f)$, due to the contribution of
induced D charges by D7 branes and the orientifold.  As in
the $SU$ case the curved worldvolume of the D7 brane induces
$-N_c/2$ units of D3 charge (accounting for the  $-N_c$ shift).
The orientifold 7-plane carries D7 charge, as is familiar,
and due to its curvature induces D3 charge
(giving rise to a net +4 shift),
as will be discussed at length below.  Similar statements
apply for $USp(N_c)$ for which the dual group is $USp(2N_f-N_c-4)$.
These results are in accord with field theory dualities
 for $SO(N_c)$ discussed
  in Refs.\seiberg\ref\intriligatorseiberg{K. Intriligator
and N. Seiberg, `Duality, monopoles, dyons, confinement and oblique
confinement in supersymmetric $SO(N_c)$ gauge theories', Nucl. Phys.
{\bf B444} (1995) 125; hep-th/9503179.},
 and field theory dualities for $USp(N_c)$ discussed in
Ref.\ref\pouliot{K. Intriligator and P. Pouliot, `Exact superpotentials,
quantum vacua and duality in supersymmetric $SP(N_c)$ gauge theories',
Phys. Lett. {\bf B353} (1995) 471; hep-th/9505006.}.
These field theory
dualities can be `derived' by flowing from the $N=2$
versions of the above gauge theories \ref\argyrespsei{P.C. Argyres, M.R.
Plesser, and N. Seiberg, `The moduli space of vacua of
$n=2$ susy QCD and duality in
$N=1$ susy QCD', Nucl. Phys. {\bf B471}(1996) 159;
hep-th/9603042.}\ref\argyrespsha{P.C.Argyres, M.R. Plesser,
and A.D. Shapere, `$N=2$ moduli spaces and $N=1$ dualities for $SO(n_c)$ and
$USp(2n_c)$ super-QCD', hep-th/9608129.}.  In fact the stringy
realization of these $N=1$, $SO$ and $USp$
dualities also starts from an $N=2$ situation,
as was the case for $SU$ \bjpsv .

Even though the action of T-duality on D-brane
charges is relatively simple, the action of T-duality
on curved
spaces with wrapped D-branes has not
been investigated before.  This
is an important subject to study further, perhaps using techniques familiar
from mirror symmetry, in order to shed further light
on field theory dualities.
 Nevertheless, the fact that $N=1$
dualities for $SU,SO, $ and $USp$ gauge theories with
fundamental matter can all be {\it derived} in a simple and unified way
suggests strongly that our assumptions about T-dualities are valid.

\newsec{The local model for $SO$ and $USp$ groups}

As mentioned in the previous section, local
models of F-theory compactified on elliptic $CY_4$'s
which admit a perturbative string description are
a natural setup to gain insight into $N=1$ dualities.
In particular for the $SO$ models we can replace
the F-theory description with a type IIB orientifold.
The type IIB orientifold description also allows
us to easily construct the $USp$ series. In this section
we  discuss the local models we need in  detail.

We start by constructing a local model for type IIB string
on a Calabi-Yau threefold.
The local model for the Calabi-Yau threefold
is the total space of a complex 2-surface $K$ together
with its canonical line bundle. This gives a non-compact
threefold with $c_1=0$.  We then consider an orientifold
of the above model by letting the space reflection part
act trivially on $K$ but as an inversion
on the line bundle over $K$.  The zero
section of the line bundle, which can be identified
with $K$, times the Minkowski space $M^4$ can be viewed as
an orientifold 7-plane, or more briefly, an O7 plane.

We now put
$N_c/2$ {\it physical} D7 branes on the O7 plane.\foot{If
the branes are moved away from the
orientifold plane, branes will have images under
the space reflection. The total number of branes would be $N_c$, out
of which $N_c/2$ branes are called physical because
their positions can be adjusted at will. While the total number of Chan-Paton
indices is $N_c$, the number of physical branes coincides
with the rank of the group.}
Depending on the choice of weight factor for string diagrams involving
crosscaps the branes
 give rise to $SO(N_c)$ or $USp(N_c)$ gauge theory.
To see this recall that in open strings we start with $N_c$ Chan-Paton
factors.  In oriented strings
this gives in the open string sector a $U(N_c)$ gauge
symmetry.
For non-orientable strings, however, we have to
 take into account the action of orientifolding on the open string
sector.  This amounts to symmetrization or anti-symmetrization of
the Chan-Paton indices, leading to $USp(N_c)$ or $SO(N_c)$
respectively.  The difference between these two
is the choice of the sign for the action of the orientifold (twist)
operator $\Omega$ on the open string sector. The two possible choices of
sign correspond, for the case of the open string one loop vacuum graphs,
to either adding or subtracting the diagrams for the annulus and the Moebius
strip. Since the Moebius strip is a disk with a crosscap,
the net effect is to weight string diagrams having no external open
strings with a weight factor $(-1)^c$ where
$c$ denotes the number of crosscaps \ref\marcussagnotti{N. Marcus
and A. Sagnotti, `Group theory from quarks at the ends of strings',
Phys. Lett. {\bf B}188 (1987) 58.}\foot{ This is a well-defined
weight factor since the number of crosscaps is well defined except for
the fact that one can trade three crosscaps for
 a single crosscap plus a handle. Thus any surface can be said
to have either two, one or no crosscaps. Only surfaces without crosscaps
are orientable.}.  We learn that to exchange
$SO$ and $USp$ all that needs to be done is to change
the sign of the contribution of the crosscap.
In particular a diagram with a single crosscap comes with opposite
signs in $SO(N)$ and $USp(N)$ theories. This fact will be used below.

If $K=T^4$ one finds $N=4$ gauge theory in four dimensions.
If $K=K3$ one finds $N=2$,
 and if $K$ is generic, one finds
$N=1$ gauge theory in four dimensions,
with $h^{2,0}(K)+h^{1,0}(K)$ matter multiplets in the adjoint of the
gauge group \katzvafa .
We will mainly consider in the latter case the situation with
$h^{2,0}=h^{1,0}=0$.  The generalization when they are non-vanishing
is exactly as in \bjpsv\ and will not be repeated here.

If we wish to obtain matter in the fundamental
representation of the gauge group of the seven branes
we have to bring in some D3 branes filling the spacetime, and put them
on $K$, where they appear as points.
This case was studied in the context of D5 branes
and D9 branes in \ref\wittensmall{E. Witten, `Small
instantons in string theory', Nucl. Phys.
{\bf B460} (1996) 541; hep-th/9511030.}\ref\gp{
E. Gimon and J. Polchinski, `Consistency conditions for orientifolds and
D-manifolds' Phys. Rev. {\bf D54} (1996) 1667; hep-th/9601038.}\ a case
equivalent to ours by T-duality.  In the case of $SO(N_c)$
 if we bring in $N_f$ D3 branes and put them all at the same
point on $K$ we obtain an $N=2$, $USp(2N_f)$ symmetry in
four dimensions, with extra
matter in the {\it antisymmetric} representation $N_f(2N_f-1)$, as well
as half-hypermultiplets in the mixed representation $(N_c,2N_f)$.
If we choose the other sign factor for diagrams with odd number of
crosscaps,
the roles of $SO$ and $USp$ are exchanged.  In particular, we get
$USp(N_c)$ from D7 branes and the extra sector we obtain
from D3 branes is an $N=2$ system
with $SO(2N_f)$ gauge symmetry.  Moreover, in addition
to the mixed matter half-hypermultiplets in $(N_c,2N_f)$,
we get matter in the
{\it symmetric} representation of $SO(2N_f)$, i.e.,
in the $N_f(2N_f+1)$
dimensional representation.

Summarizing, the two gauge groups we have engineered are
\eqn\gaguge{  \eqalign {
& SO(N_c)\times \,
USp(2N_f)\,, \cr
& USp(N_c)\times SO(2N_f) \,,\cr } }
and the matter content is given as follows.  For $K=T^4$, in $N=2$
hypermultiplets
\eqn\mcontttt{ \half \cdot [ N_c\, ,\, 2N_f]\,\,+\,\, [ 1 \,, \, N_f(2N_f\mp 1)
]
\, \,+ \,\, [ N_c (N_c\mp 1)/2 \,, \, 1 ] \,,}
where the top sign refers to $SO(N_c)\times USp(2N_f)$ and the lower
sign refers to $USp(N_c)\times SO(2N_f)$. The last $N=2$ hypermultiplet
arises by noting that the $N=4$ theory has an extra
adjoint hypermultiplet in the $N=2$ terminology.
For $K= K3$, we find in $N=2$ hypermultiplets
\eqn\mcontke{ \half \cdot [ N_c\, ,\, 2N_f]\,\,
+\,\, [ 1 \,,\, N_f(2N_f\mp 1) ]\,, }
i.e. the adjoint is lost due to having less supersymmetry
preserved on the D7 worldvolume.
For $K=S$, in terms of four-dimensional $N=1$
chiral multiplets we get
\eqn\mcontsf{ 2\cdot \half\cdot
 [ N_c\, ,\, 2N_f]\,\,+\,\,2\cdot \, [ 1 \,, \, N_f(2N_f\mp 1) ]\,
+ \,[1\,,\, N_f (2N_f \pm 1) ]\,, }
where the factors of 2 arise from writing an $N=2$ hypermultiplet
as two $N=1$ chiral multiplets, and the last representation
arises from writing the $N=2$ vector multiplet on D3 branes in terms
of an adjoint $N=1$ chiral multiplet.

\newsec{Matching of Higgs branches to Instanton moduli spaces}

For concreteness let us first consider the case with $SO(N_c)$ gauge group for
the D7 branes. The Higgs phase of the gauge theory models arise
in the local string models as follows. The $N_f$ D3 branes, living inside the
D7 branes, fill four dimensional Minkowski space and appear as $N_f$
points in $K$. When these $N_f$ points coincide the
$USp(2N_f)$ symmetry is unbroken and the D3 branes
can be viewed as $N_f$ coinciding
zero size instantons of $SO(N_c)$. Each three brane
contributes an instanton number of $+1$, giving a total instanton number
$k=N_f$. In this section we test our construction of local models
by checking that the dimensionality of the Higgs branches of
the field theories match the dimensionality of the corresponding
moduli space of instantons on
 $K$ in the local string model. This can be viewed as
an extension of Ref.\wittensmall.  The fact
that not only the dimension but also
the local structure of the instanton moduli space
matches the Higgs description
follows from \wittensmall\ for the $N=2$ case.  For the
$N=1$ case presumably an extension of the $N=2$
result works; such extension was
established for $N=1$ supersymmetric $SU(N)$ gauge
theories in
\bjpsv .

\medskip
We begin by recalling the
dimensionalities of moduli spaces of instantons.
Consider a gauge theory based on the gauge group $G$ and living on a four
dimensional manifold $K$. The moduli space ${\cal M}_k(G, K)$
of instantons with total instanton number $k$ has complex dimension
\eqn\dimm{\half \hbox{dim}_{{}_{\bf C}} {\cal M}_k (G, K) =
 k  c_2(G) \,- {\textstyle {1\over 8}} \,[\chi (K)
+ \sigma (K) ]\,\,\hbox{dim} (G) \,, }
where $c_2(G)$ is the dual Coxeter number of the group $G$ and
$\chi (K)$ and $\sigma (K)$ denote the Euler characteristic
and the signature of $K$ respectively. As mentioned earlier, we have
three cases of interest: $K= T^4, K3$ and $ S$. For $T^4$ we have $\chi +
\sigma=0$,
for $K3$ we have $\chi + \sigma = 8$, and for $S$ we have $\chi + \sigma =4$.
We therefore write
\eqn\thedims{\eqalign{
\half \hbox{dim}_{{}_{\bf C}} {\cal M}_k (G, T^4) &=
 k  c_2(G) \,,  \cr
\half \hbox{dim}_{{}_{\bf C}} {\cal M}_k (G, K3) &=
 k  c_2(G) \,- \,\hbox{dim} (G)\,, \cr
\half \hbox{dim}_{{}_{\bf C}} {\cal M}_k (G, S) &=
 k  c_2(G) \,- \half\,\hbox{dim} (G) \,.\cr}  }
We also list, for easy reference, the dual Coxeter numbers of the groups we
will be dealing with
\eqn\coxeter{c_2(SU(N)) =N\,,\quad c_2(SO(N)) = N-2\,,\quad c_2(USp(N))= \half
N +1\,.}
\medskip

We can now consider the Higgs branches of the supersymmetric field theories.
In $N=2$ four dimensional gauge theory we count the (real) dimensionality
of the Higgs branch ${\cal M}_H$ using the equation
\eqn\dimhiggs{{\textstyle {1\over 4}}
 \hbox{dim}_{{}_{\bf R}} {\cal M}_H = \# \hbox{hyp.}-
\hbox{dim} (G)\, . }
Consider first the case of $T^4$, which gives rise to the gauge group
$SO(N_c)\times USp(2N_f)$, with the matter representations listed in
\mcontttt.  We find
\eqn\dimhiggs{\eqalign{
{\textstyle {1\over 4}} \hbox{dim}_{{}_{\bf R}} {\cal M}_H &=
N_cN_f + N_f (2N_f-1) + \half N_c(N_c-1) \cr
&\quad\quad  -  \half N_c(N_c-1) - N_f (2N_f +1) \, ,\cr
&= (N_c-2) N_f \, ,\cr }  }
in  agreement with the expectation from the first equation in \thedims,
for the case of $G=SO(N_c)$. Had the $USp(2N_f)$ part of the gauge symmetry
been completely broken down to $[USp(2)]^{N_f}$
by pulling the D3 branes apart on $K$, the dimensionality
of the Higgs branch in the resulting theory would have still
been the same.

For the case when $K=K3$ we still have an $N=2$ gauge field theory. We must
now use the matter representations listed in \mcontke\ which now give
\eqn\dimhiggske{\eqalign{
{\textstyle {1\over 4}} \hbox{dim}_{{}_{\bf R}} {\cal M}_H &=
N_cN_f + N_f (2N_f-1)  -  \hbox{dim} SO(N_c) - N_f (2N_f +1)\,, \cr
&= (N_c-2) N_f - \hbox{dim} SO(N_c) \,, \cr }  }
in  agreement with the expectation from the second equation in \thedims,
for the case of $G=SO(N_c)$.

Finally, we consider the case when $K= S$.  Let us first
note that the last representation listed in \mcontsf\ which
arises from an $N=2$  vector multiplet comes with a superpotential
term $W$ induced from $N=2$ supersymmetry.  Solving the $dW=0$ constraint
implies that this field appears as a constraint on the Higgs branch
and thus acts to decrease the dimension of Higgs branch, rather
than add to it.  We thus find (counting in chiral multiplets
which correspond to computing ${1\over 2}{\rm dim}_{\bf R}{\cal M}_H$)
\eqn\dimhiggsone{
\eqalign{ \half \hbox{dim}_{{}_{\bf R}} {\cal M}_H
&= 2N_fN_c + 2N_f (2N_f-1) - N_f (2N_f +1) \cr
&\quad\quad - \hbox{dim} (USp(2N_f)) -\hbox{dim} (SO(N_c)) \,,\cr
&= 2N_f (N_c-2) - \hbox{dim} (SO(N_c)) \,,} }
in  agreement with the third equation in \thedims,
for the case of $G=SO(N_c)$ (note that one equation refers to complex
dimension while the other refers to real dimension).

\medskip

For the D7 branes giving $USp$ the story is similar.
The main difference is that now the instanton number
for $USp$ is related to $N_f$ by $k=2N_f$. This is in agreement
with \ref\douglas{M. Douglas, `Branes within branes', hep-th/9512077.}
where it was explained that the $k$-instanton moduli space of
$USp$ theories is governed by an $SO(k)$ gauge theory. Indeed, in
our case the $SO$ group induced by the three branes is $SO(2N_f)$.
To understand this counting
explicitly
note that we can have integral or half-integral
D-brane charges for the $SO$ group. Only pairs
of half-integral D-branes can be moved off the orientifold
plane.  In the case we have been considering we have
brought D3 branes onto the orientifold plane O7, and thus
each one will count as two instantons of zero size on top of
each other.  Once they are on O7 they can split into two
half-integral D3 charges, each of which would correspond
to one  zero-size instanton of $USp(N_c)$.   Note in particular
that for one instanton of zero size in $USp(N_c)$ we would get
an $SO(1)$ group corresponding to a half D3 brane stuck on
the orientifold plane. The absence of
gauge group in this case implies the absence of a
scalar degree of freedom which can move
the D3 brane off the orientifold plane.
 We thus conclude
that for $N_f$ D3 branes the instanton number
for $USp(N_c)$ is $k=2N_f$.  It is now straightforward
to repeat the above analysis for the $USp$ D7
branes and show that the dimensions
for the instanton moduli spaces match the  dimensions
of the Higgs branches.

\newsec{R-charge and Instanton Corrections}

It was shown in \bjpsv\ that point instantons
of $SU(N_c)$ supersymmetric QCD can be identified
with Euclidean D3 branes wrapped around $S$.  Moreover,
following \ref\wimfin{E. Witten, `Non-perturbative superpotentials
in string theory', hep-th/9604030.}\
an R-charge $Q$ was defined so that
instantons contribute to the superpotential if there is
a charge violation $\Delta Q = 1$.
In this section we will extend these considerations to the
local models for $SO$ and $USp$ gauge theories. With this end
in mind, let us first review the calculation of charge violation
in the $SU(N_c)$ local model of \bjpsv.

We have an Euclidean D3 brane wrapping
$S$ and we assume we are in the Higgs branch, namely, there are
$SU(N_c)$ background gauge fields on $S$ of instanton number $N_f$.
The total charge violation arises from additive contributions
from two sectors. The first sector, arising from open strings
stretched from the D7 branes to
the Euclidean D3 brane, represents
matter in the fundamental of $SU(N_c)$, living
on $S$ and interacting with the background
$SU(N_c)$ gauge fields.  The charge violation
in this sector is given by the
index, in the fundamental representation,  of the twisted
Dirac operator $\overline \partial_A$,
with $A$ denoting the $SU(N_c)$ background gauge fields.
This index is given by
\eqn\numb{\Delta Q|_{D3- D7}=
{\rm index}{\ \overline \partial}_A=N_c-N_f.}
The second sector arises from open strings stretched from the Euclidean D3
brane onto itself, and corresponds to a twisted
$N=4$ supersymmetric
$U(1)$ theory  on the Euclidean D3 brane.
Certain aspects of the field theory on this Euclidean
instanton have been studied recently in \ref\gan{O. Ganor,
`A note on zeroes of superpotentials in F-theory', hep-th/9612077.}.
The contribution to $\Delta Q$ from this sector
is zero, because the fermions in the hypermultiplet
and vector multiplet carry opposite $Q$ charge\foot{This follows
because the twisting of $N=4$  is related
to the $Q$ charge.  In particular four scalars
in the hypermultiplet as well as the gauge field of
the $N=4$ are neutral under $Q$ which forces the fermions in the vector
and hypermultiplet to carry opposite $Q$ charge.}
and they have equal number of zero modes, i.e. from this sector we have
\eqn\acj{\Delta Q|_{D3- D3}=\# {\rm hyp.} -\# {\rm vect.}=0\,.}
The total charge violation is thus given by $\Delta Q=N_c-N_f$.
One expects superpotential generation for $\Delta Q =1$, or equivalently,
for $N_f = N_c-1$. This expectation is confirmed
by field theory analysis.

Now we come to the case
of $SO$ and $USp$ gauge theories.  Let us consider the $SO$
case first. The $SO(N_c)$ background gauge fields on $S$ define a background
with  instanton  number $N_f$, and we have
an Euclidean D3 brane wrapped around $S$.
Once again, we have two sectors contributing to charge violation.
{}From D3$-$D3 open strings, the Euclidean D3 brane
will carry an $N=2$ supersymmetric
$USp(2)$ gauge theory with an additional singlet hypermultiplet.
This sector gives a contribution
\eqn\socont{
\Delta Q|_{D3- D3}=\# {\rm hyp.}-\# {\rm vect.}=1-3=-2\,.}
{}From D3$-$D7 open strings, we have a
half-hypermultiplet
in the $[2,N_c]$ representation of $USp(2)\times SO(N_c)$.
As before,
the violation of $Q$ charge in this sector is given by
the index, in the $[2,N_c]$ representation, of ${\overline
\partial}_A$, where $A$ is the nontrivial $SO(N_c)$ background.
We find
\eqn\secondrb{\Delta Q|_{D3- D7}=N_c-2N_f\,. }
We thus find that the total violation is given by
\eqn\totvioso{\Delta Q=N_c-2N_f-2\,. }
We thus expect that for $\Delta Q=1$, i.e. when the number of
flavors $2N_f$ is given by  $2N_f = N_c- 3$  we have a superpotential
generated.  Moreover, as in \bjpsv,
one can see that the superpotential must have a first order pole.
This is in accord with field theory expectations \intriligatorseiberg .
The situation, however, is more complicated than
in the $SU$ case,  because for $\Delta Q=1$
we now have an unbroken $SO(3)$ gauge theory with non-trivial
infrared dynamics and gaugino condensates will play
a role in the creation of the superpotential.  Similarly,
for $\Delta Q=0$, corresponding to $2N_f = N_c-2$, there
could be a quantum correction to the moduli space.  As it turns out,
in this case there are instanton corrections to the coupling
of the unbroken $U(1)$ on the moduli space.
These have been computed exactly in \intriligatorseiberg\ and it would be
interesting
to see how the above Euclidean D3 instanton (and multi-instantons)
reproduce those results.\foot{It would be interesting to see what
happens if we turn on $USp(2)$ instantons
on the Euclidean D3 brane
and sum over all possible values of instanton number.}

The situation for the $USp(N_c)$ gauge group is similar, modulo
a small twist, as in the previous section.  If we consider
one  Euclidean D3 brane wrapped around $S$, as
noted in the previous
section, it corresponds to {\it two} zero-size
instantons on top of each other.
Since this is most similar to the case considered above
let us first compute
the $Q$ charge violation for two instantons and at the end
divide the
result by two
to obtain the charge violation for {\it one}
Euclidean D3 instanton.  For two instantons
 we have an $SO(2)$ gauge group living on the Euclidean
D3 brane. The R-charge violation from the mixed sector
is the same as before.  But the violation from the D3-D3
sector is different, because now we have three hypermultiplets
(symmetric representation of $SO(2)$) and one vector multiplet
(adjoint of $SO(2)$) and we find
\eqn\chvi{2\Delta Q|_{D3-D3}=3-1=2\, .}
Thus the net violation of the $Q$ charge is
\eqn\anchr{2\Delta Q= N_c-2N_f+2\quad \to \quad \Delta Q= \half N_c-
N_f +1\, .}
Note that we could have also done the computation
directly in terms of a {\it single} instanton corresponding to
an  Euclidean half D3 brane. This
leads to an $SO(1)$ theory on the Euclidean brane
with one hypermultiplet in the D3$-$D3
sector, and no vector multiplet.
This gives the contribution $1-0=1$ to the charge violation.
The mixed D3$-$D7 sector will give a single half-hypermultiplet
contributing $\half N_c-N_f$ charge violation. These two contributions
add up to the result in \anchr.

For $\Delta Q=1$ we thus expect, as before,
a superpotential with a first order pole be generated by point-like
instantons.  This corresponds to the case $\half N_c=N_f$, where indeed
there is a superpotential generated with a first
order pole \pouliot.\foot{In comparing with \pouliot\ note that the authors
work with $SP(N_c)$, a group that in our notation is $USp(2N_c)$. The
number of flavors, $2N_f$, is the same as in our case. To compare
with our results simply let $N_c\to N_c/2$ in the results of \pouliot.}
We  would also expect that for $\Delta Q=0$ there could be
an instanton correction to the moduli space.  This is indeed
the case \pouliot.

\newsec{$N=1$ duality as T-duality}

In this section,
as a useful preliminary to the analysis of $N=1$ supersymmetric
$SO$ and $USp$ gauge theories we first  review the results of \bjpsv\ dealing
with $SU$ gauge theories. We then turn to the computation of induced
charges by orientifold planes  and D-branes, and finally derive the
dualities of $N=1$ supersymmetric
$SO$ and $USp$ theories from T-dualities of the local models.

\subsec{Review of the SU case}

In this case  one
starts with $K=K3$ with $N_c$ D7 branes wrapped on it
and $N_f$ D3 branes appearing as points on it.
The $N_c$ D7 branes wrapped on $K3$
induce  $-N_c$ units of D3 brane charge \ref\bsv{M. Bershadsky, V. Sadov and
C. Vafa, `D-branes and topological field theory', Nucl. Phys. {\bf B463}
(1996) 420; hep-th/9511222.}
\ref\mgh{M. Green, J. Harvey and G. Moore, `I-brane inflow and anomalous
couplings of D-branes', hep-th/9605033.}.
For later application let us note that, by T-duality, the above statement
implies that a D-brane wrapped on $K3$
gives rise to a codimension-four D-brane charge of $(-1)$.
 Thus in this configuration the total seven brane charge $Q_7$
and the total three-brane charge $Q_3$ are given by
\eqn\fistcase{
\eqalign{ Q_7 &=N_c\,, \cr
Q_3 &=N_f-N_c\,. \cr}  }
If the volume of $K3$ is small we are
in a regime which by T-duality
is equivalent to a large volume $K3$.  Note that perturbatively
the volume of $K3$ is related to the bare coupling of the D7 gauge
group $SU(N_c)$ by $V(K3)=1/g^2$. Thus, when we enter the
small volume regime we are entering a large coupling region of
the original theory.  Once we dualize we are back to a weak
coupling description.
This, however, exchanges the D3 and D7 charges.  In particular now we
have
\eqn\newch{\eqalign{
Q_7'&= Q_3 =N_c'\,, \cr
Q_3' &=  Q_7= N_f' - N_c' \,. \cr} }
Solving for $N_c'$ and $N_f'$ we find
\eqn\newchi{\eqalign{
N_c'&=  N_f-N_c \,, \cr
N_f' &=  N_f \,. \cr} }
This means we now have a weak coupling description of $SU(N_f-N_c)$
with $N_f$ flavors in the fundamental
of this group.  This is still an $N=2$ theory. To obtain
an $N=1$ theory we assume $K3$ has an extra $Z_2$ symmetry
which inverts the sign of the holomorphic 2-form.  We can now mod
out the $K3$ by this $Z_2$ symmetry to obtain a space
$S=K3/Z_2$ which has $h^{2,0}=0$.  Note that this $Z_2$ acts only
on the middle cohomologies of $K3$.
This does not interfere with the
D7 and D3 brane charges as they come from zero and four dimensional
cycles of $K3$.  So we expect the same duality to continue to hold
by the time we get to $N=1$.  This is very much in the same
spirit as the flow from $N=2$ microscopic/macroscopic theories
to the dual pairs of $N=1$ theories \argyrespsei .
Moreover, the meson field of the dual magnetic theory can naturally
arise if the $N_f$ D3 branes in the dual theory are forced to be
on top of each other.

\subsec{Computation of induced charges for orientifold planes and D-branes}

We now wish to repeat the same count of D-brane
charge for the $SO$ and $USp$ theories, and see if they
are in accord with the expectations.  Again we start with
$K=K3$ and assume that modding $K3$ by the extra $Z_2$
will not modify the D-brane charge count.  This would then
be in the same spirit as the flow from the $N=2$ microscopic/macroscopic
theories to the dual $N=1$ theories \argyrespsha .
We will assume that the D3 branes are separate.   In the
case of $SO$ D7 branes this implies that we have
$USp(2)^{N_f}$.  But the dynamics of these gauge factors
will be infrared trivial if $N_c>4$, which we will be assuming.  Therefore,
these factors will not affect the field
theory dualities.  A similar comment applies to the
case of $USp$ D7 branes.

The main complication for the $SO$ and $USp$ cases
 compared to the $SU$ case is that, in addition to the D7 branes, the
orientifold planes can also induce
D-brane charges.  The D7 and O7 induce 3-brane charges
through an interaction of the form $\int B^4 \wedge tr (R \wedge R)$
where $B^4$ is the gauge field coupling to D3 charges.   This arises
for the D7 case from a disk diagram with the boundary on the
D7 branes.  For the O7 this term arises from the sphere
with a single crosscap ($RP^2$).  That these are generated
follows from the anomaly cancellation considerations for
type I strings (which by T-duality is related to the above
interactions).  In addition, there is the familiar D7 charge
induced by the O7. Let us study these contributions in detail.

\noindent
\underbar{D3-brane charge induced by curved D7-branes} ~Each
physical D7 brane will contribute some $Q_3$ charge. This
is because the worldvolume of the seven brane is curved
and the curvature
of $K3$ is responsible for generating an effective D3 brane charge,
as in the $SU$ case.
As mentioned above the string diagram computation involves
a disk worldsheet.
Being an orientable diagram it is also present
with exactly the same value in the non-orientable case we are now
dealing with.   Thus, as before, each D7 brane will
contribute a $Q_3$ charge
\eqn\dthrindn{Q_3 (\hbox{D7}) = -1 \,,}
irrespective of whether we are dealing with $SO$ or $USp$ seven branes.
Note that for $N_c/2$ D7 branes we will end up getting $-N_c/2$ units of
D3 charge.
\medskip

\noindent
\underbar{D7-brane charge of O7 planes}
The contribution of the O7 plane to the seven brane  charge
$Q_7$ is easily determined once
we note that O9 plane  carries $(-16)$ units of D9 charge
(the explanation of $SO(32)$ gauge symmetry
as orientifold of type IIB in 10 dimensions \ref\pol{J. Polchinski,
`Tasi Lectures on D-branes', hep-th/9611050.}).
Once we compactify on a two-torus $T^2$ and do T-duality we obtain
four O7 planes with the same total charge as before, namely $(-16)$
units of D7 charge.
Thus each O7 plane carries $(-4)$ units of seven brane charge $Q_7$.
Note that since the
contribution of the O7 plane to D7 charge comes from a diagram
with a single crosscap ($RP^2$) and going from $SO$ to $USp$ involves
the change in sign for diagrams with odd number of crosscaps,
the O7 plane will carry $+4$ units of seven brane charge $Q_7$
for the $USp$ case. Thus we write
\eqn\osevench{Q_7 (\hbox{O7}) = \mp 4 \,,}
where the top sign refers to $SO$ D7 branes
and the lower sign refers to $USp$ D7 branes.

\medskip
\noindent
\underbar{D3-brane charge of curved O7 planes}~
 We now wish to compute
the D3 charge induced by an O7 wrapped around $K3$.
As noted above, this comes from an $RP^2$ diagram
producing an interaction of the form $\int B^{4}\wedge tr (R\wedge R)$.
To get the normalization, it is convenient to proceed as follows.
 Let us
first compute the D5 charge of an O9 wrapped around $K3$.
In the
ten dimensional type I theory compactified on $K3$
the $16$ D9 branes and the single O9 plane both
contribute to the five brane charge $Q_5$.
As is familiar \wittensmall ,
upon $K3$ compactification
in order to cancel the D5 charge induced by
O9 and D9 we require the
addition of $24$ five-branes. This means that the induced charges satisfy
\eqn\constr{ 16\cdot Q_5 (\hbox{D9}) + Q_5 (\hbox{O9}) = -24 \,. }
 It follows from our earlier comments
that $Q_5(\hbox{D9})=-1$, and
therefore from \constr\ we find that $Q_5(\hbox{O9}) = -8$.

To find the D3 brane charge of O7 planes
wrapped on $K3$ we use the same trick as before, namely compactify
on $T^2$ and dualize.  In this process we obtain four O7 planes
wrapped on $K3$, and therefore each O7
plane wrapped on $K3$ will induce $-8/4=-2$ units of D3
brane charge.  Note that this is for the case of $SO$ gauge groups.
When we deal with the $USp$ groups, as noted earlier,
the contribution of the diagrams with a single
crosscap change.  This implies that for the $USp$ case the
induced D3 brane charge is $+2$ units. All in all we find
\eqn\threebrcha{Q_3 (\hbox{O7}) = \mp 2 \,, }
where the top sign refers to $SO$
and the lower sign refers to $USp$.

\medskip
\noindent
\underbar{Total three brane and seven brane charges}
In the local models we have been
considering we have focused on a single orientifold O7 plane and have placed
a total of $N_c/2$ physical D7 branes on top of it giving
rise to the gauge group $ SO(N_c)$ or $USp(N_c)$.
It follows that the total amount $Q_7$
of D7 brane charge of this local configuration is given by
the number of branes plus the orientifold contribution \osevench\
\eqn\dscharge{Q_7 = \half N_c \mp 4 , }
where the top and lower signs in the above refer to
$SO(N_c)$ and $USp(N_c)$ respectively.

The total three brane charge $Q_3$ arises
from the $N_f$ physical $D3$ branes, the D7 brane contribution
indicated in \dthrindn\ multiplied by the number of seven branes
$\half N_c$,  and the O7-plane contribution
indicated in \threebrcha\
\eqn\threebch{Q_3 = N_f - \half N_c \mp 2 \,.}

\subsec{T-Duality transformations for the $SO$ and $USp$ local models}

We are now
ready to see the effect of T-duality on the
above charge assignments.  First of all we need to
argue that T-duality brings the $K3$ orientifold back
to itself.  This would {\it not} be the case if
we were dealing with $T^4$ where $T$-duality would have
turned the O7 plane into O3 planes.  To see that it is reasonable
for $K3$ note that, if  we realize $K3$ as $T^4/Z_2$
the orbifold group is generated by the elements
 $\{ 1,\Omega, I, I\Omega \}$ where
$\Omega$ refers to the orientifold action and $I$ refers to the
$Z_2$ inversion of $T^4$.  Under T-duality, $I$ goes to itself
whereas $\Omega$ and $I\Omega$ get interchanged, so we
end up again with a $K3$ orientifold.  This argument leads us
to believe that T-duality will bring the $K3$ orientifold
back to itself even away from the orbifold
limit of $K3$ (which we need to assume in order to avoid the zero size
instantons
concentrated at the fixed points of the orbifold,
as discussed in \ref\berkooz{M. Berkooz, R. Leigh, J. Polchinski,
J. Schwarz, N. Seiberg, E. Witten, `Anomalies, dualities, and topology
of $D=6$, $N=1$ superstring vacua', hep-th/9605184.}).
At any rate we shall assume that for a smooth $K3$ orientifold the
T-dual is still a $K3$ orientifold.

Now let us concentrate on the total D-brane charges in our
local models. These were given in eqns.\dscharge\ and \threebch,
which we reproduce for convenience here
\eqn\harge{\eqalign{
Q_7 &= \half N_c \mp  4 \, ,\cr
Q_3& =N_f - \half N_c \mp 2 \,.\cr} }
Since T-duality exchanges the D3 and D7 charges, in the dual
model, the seven brane charge $Q_7'$ and the three brane charge
$Q_3'$ are given by
\eqn\arge{\eqalign{
Q_7' &= Q_3 = \half N_c' \mp 4 \, , \cr
Q_3'& = Q_7 = N_f' - \half N_c' \mp 2 \,,\cr} }
where $N_c'$ and $N_f'$ denote respectively
the number of D7 branes and D3 branes
in the dual model. Solving for $N_c'$ and $N_f'$ we find
\eqn\arge{\eqalign{
N_c'  &= 2N_f - N_c \pm 4 \, ,\cr
N_f'& = N_f\,.\cr} }
This implies that under a T-duality transformation the gauge groups in the
local models get exchanged as
\eqn\dualexch{\eqalign{
&SO(N_c)\,\,\leftrightarrow \ SO(2N_f-N_c+4)\,,\cr
&USp(N_c)\leftrightarrow USp(2N_f-N_c-4)\,,\cr}}
each theory with $2N_f$ flavors in the fundamental representation
of the corresponding group.  These are the familiar $N=1$ dualities
for the $SO$ and $USp$ groups with matter in the fundamental
\intriligatorseiberg\pouliot.
The additional fundamental meson fields appearing in the magnetic side may
arise in the local models (as in \bjpsv) if in the T-dual theory
the $N_f$ D3 branes are forced to be near each other.
In this case, the superpotential term involving the fundamental
meson field on the magnetic side will also arise as
expected.

\bigskip
We would like to thank M. Bershadsky, P. Cho,
 K. Intriligator, A. Johansen, T. Pantev and V. Sadov
for valuable discussions.

\bigskip
The research of CV is supported in part by NSF grant PHY-92-18167.
The research of BZ is supported by D.O.E. Cooperative grant
DE-FC02-94ER40818, and a grant from the John Simon Guggenheim Foundation.

\listrefs
\end